# GENAI AS DIGITAL PLASTIC: UNDERSTANDING SYNTHETIC MEDIA THROUGH CRITICAL AI LITERACY

Jasper Roe [1*], Leon Furze [2], Mike Perkins [3]

[1] Durham University, United Kingdom
[2] Deakin University, Australia
[3] British University Vietnam, Vietnam

[*] Corresponding Author: jasper.j.roe@durham.ac.uk

February 2025

## Abstract

This paper introduces the conceptual metaphor of 'digital plastic' as a framework for understanding the implications of Generative Artificial Intelligence (GenAI) content through a multiliteracies lens, drawing parallels with the properties of physical plastic. Similar to its physical counterpart, GenAI content offers possibilities for content creation and accessibility while potentially contributing to digital pollution and ecosystem degradation. Drawing on multiliteracies theory and Conceptual Metaphor Theory, we argue that Critical Artificial Intelligence Literacy (CAIL) must be integrated into educational frameworks to help learners navigate this synthetic media landscape.

We examine how GenAI can simultaneously lower the barriers to creative and academic production while threatening to degrade digital ecosystems through misinformation, bias, and algorithmic homogenization. The digital plastic metaphor provides a theoretical foundation for understanding both the affordances and challenges of GenAI, particularly in educational contexts, where issues of equity and access remain paramount. Our analysis concludes that cultivating CAIL through a multiliteracies lens is vital for ensuring the equitable development of critical competencies across geographical and cultural contexts, especially for those disproportionately vulnerable to GenAI's increasingly disruptive effects worldwide.

# Introduction

Generative Artificial Intelligence (GenAI) has changed the nature of writing through its ability to craft human-like, convincing text and other content, including the multimodal production of video, audio, images, and digital artifacts. Producing computer code, creating applications, and supporting research activities are all now within the scope of these new technologies (Perkins & Roe, 2024). GenAI agents are poised to produce novel data, complete tasks, and automate aspects of creative processes without guidance or supervision, leading to unknown consequences for society and educational contexts. This rapid progression has taken place in an extremely narrow temporal window, and resonates with the New London Group's (1996) conceptualization of multiliteracies, in which traditional methods of language learning are insufficient to account for the multiple methods of creating meaning (Lim & Querol-Julián, 2024).

Consequently, the multiliteracies project is more relevant than ever as a result of both GenAI and the proliferation of multimodal texts and media we are exposed to in modern digital environments (Lim, 2024), such as those on social media networking sites, including YouTube and TikTok (Lim & Querol-Julián, 2024). Today's technology users constantly interact with a vast array of different forms of multimodal media, and increasingly such media may be synthetic in origin. Questionable content arising from actors who proliferate disinformation, online trolls, and influencers in the quest for interactivity, along with traditional media producers, competes for views in the digital attention economy.

As a result of this proliferation of multimodal and synthetic content, today's users of digital tools must develop new literacies to allow them to engage sensitively with these forms of meaning. The recent production of AI-generated Instagram and Facebook profiles and the subsequent backlash against harmful stereotypes (Bhuiyan, 2025) demonstrate the confusing and evolving nature of this situation. Deepfake technologies are poised to make it increasingly difficult for participants in the digital world to know whether what they are seeing is real or an AI-generated imitation of reality, as technology can now effectively portray people doing and saying things that they have never done (Ferrara, 2024; Roe, Perkins, & Furze, 2024). In pedagogy and assessment, GenAI tools may render traditional approaches useless (Kalantzis & Cope, 2025), and consequently, frameworks to evolve assessment in education have become a topic of conversation (Furze et al., 2024; Liu & Bridgeman, 2023; Perkins, Furze, et al., 2024; Perkins, Roe, & Furze, 2024; Roe, Perkins, & Tregubova, 2024).

To adapt to this changing climate, the core tenets of multiliteracies remain, yet may now require expansion to include new forms of AI literacy. We contend that within this space, Critical Artificial Intelligence Literacy (CAIL) (Roe, Furze, & Perkins, 2024) as one aspect of a broader multiliteracies pedagogy, is a vital competency to ensure that individuals are ready for this brave new world, especially the younger generations who will live, work, and learn within this environment. Simultaneously, existing inequalities and entrenched differentiators of epistemic capital continue to emerge in new, AI-mediated forms, and the development of such CAIL as part of multiliteracies must remain committed to educational justice.

We define CAIL as the ability to understand, appraise, and evaluate synthetic, AI-produced text and media. This includes being able to notice and evaluate GenAI content 'in the wild', as GenAI is leaving an imprint on written language through the use of 'telltale adjectives' (Singh

Chawla, 2024) or 'footprints' (Tang & Eaton, 2024). The identifiably sycophantic (Sharma et al., 2023) outputs of GenAI models such as ChatGPT are making their way into multiple online contexts, and flooding online spaces with 'AI slop' (Hoffman, 2024), yet we may be unprepared to identify these cases, especially as AI text detection has been shown to be fallible and unreliable (Perkins, Roe, Postma, et al., 2024; Perkins, Roe, Vu, et al., 2024; Weber-Wulff et al., 2023) and with the ever-increasing technological improvements of these mod. In scientific research and publication, a significant proportion of researchers use GenAI to assist with writing papers (Bjelobaba et al., 2024; Singh Chawla, 2024). One study suggests that close to 1% of all scholarly articles published in 2023 contained some evidence of GenAI use (Gray, 2024), demonstrating the intensity and rapidity with which AI-generated text is becoming commonplace in academic and scholarly environments.

While the proliferation of GenAI in academic writing exemplifies its growing influence, it represents only one aspect of a broader transformation in digital content creation. Crucially, these AI-generated outputs are fundamentally created from human-created data, scraped from the Internet without owner consent, leading to new conversations about copyright and the 'theft of collective intelligence' (Kalantzis & Cope, 2025) for training purposes. Research also suggests that human-created data is fundamental to model development, as training a new model on existing GenAI content can lead to model collapse, making models produce unintelligible and nonsensical output. This has been labelled pollution of the dataset (Shumailov et al., 2024), and we draw on this concept in this paper when mobilizing a new metaphor of 'digital plastic' as a means to foster CAIL.

Throughout the following sections, we attempt to highlight the properties of synthetic, AI-produced media to foster CAIL as part of a new aspect of the multiliteracies project. This is achieved by introducing the theoretically grounded conceptual metaphor of 'digital plastic', which draws on the properties of physical plastics to demonstrate the myriad properties of synthetic media from a multiliteracies perspective. A key part of the digital plastic metaphor that is compelling is its paradoxical nature and inability to fit into a dichotomy of 'good and bad.' Rather, the consequences of GenAI content must be viewed as complex, shifting, and contextual. Just as physical plastics revolutionized consumer society because of their flexibility, durability, and cost-effectiveness, synthetic media offers a similar set of features. Rapid, scalable, multimodal content creation can be undertaken within seconds, tailored to use cases, and often at no cost to the end user. Simultaneously, these properties may degrade the integrity of our digital worlds, just as plastic degrades the quality of the natural environment and pollutes the ecosystem. The notion of a 'monoculture' of knowledge is relevant here (Messeri & Crockett, 2024), which draws on a metaphor for the planting and cultivation of a single crop at the expense of production methodologies that, owing to their variability, are more resistant to unexpected environmental or economic changes. In the same way, it can be argued that the knowledge-effects of GenAI models reproduces dominant ideologies and cultural values (Roe, 2024), yet at the same time producing output that is less varied and inclusive of other forms of knowledge or knowing. However, we argue that plastic is a more apt interpretation for developing pedagogies that seek to enhance CAIL.

The metaphor of digital plastic invites us to rethink literacy in an age that is increasingly dominated by GenAI. Synthetic media, generated by GenAI, disrupts our understanding of literacy by collapsing the existing strata (Kress & Van Leeuwen, 2001) of meaning-making and reconfiguring the relationships between creators, technology, and audiences. Against this backdrop of intensified, digitalized multimodal meaning-making (Lim, 2024), critical theories of multiliteracies can be used to develop the next generation (Mirra et al., 2018) and enable them to tease apart and analyze the effects of GenAI content. By integrating the concept of digital plastic with multiliteracies theory, we propose a new interpretation for understanding the implications of synthetic media. This begins by examining a core tenet of multiliteracies as formulated by the NLG: educational justice.

## Multiliteracies, GenAI, and Educational Justice

At the core of the NLG's multiliteracies project is a dedication to improving equity and social and educational justice, attempting to mediate the exclusion of marginalized groups along ethnic and cultural lines (Garcia et al., 2018). In this case, the advent of GenAI may pose a fundamental threat to the core value of the multiliteracies agenda. Leading voices in the field have identified that the increasingly available, powerful AI technologies that may benefit learning may mainly be realized by users in countries in the Global North. Consequently, this disparity may worsen the digital divide and magnify entrenched educational inequalities (Miao & Holmes, 2023). Even within the same country or geographical context, access to technologies that benefit literacy for learners is variable among students (Chandel & Lim, 2024), which could further magnify inequalities. To demonstrate further, those without access to paid subscription models, high-speed Internet, or applications and technologies required to benefit from GenAI may find themselves further disadvantaged, while those who do have access to these forms of cultural, social, and economic capital will be able to reap outsized benefits.

There are multiple ways in which GenAI can drive inequalities. Companies that develop such models tend to be well-funded, private corporations from the Global North, which use labor from countries in the Global South to help train the models and simultaneously engage in market capture (Zhgenti & Holmes, 2023). Such an approach means that the potential for countries in these geographical regions to develop their own models is limited, especially as training and producing AI models is highly resource-intensive and relies on specialized technology such as Graphical Processing Units (GPUs). This lack of locally produced AI systems leads to a reliance on models and tools produced by large corporations, which have been trained mainly on Western-centric and English-language databases. This situation creates a potential for technological dependency while also compelling engagement with systems that perpetuate algorithmic and cultural biases, thereby undermining the educational justice. This differing range of access may drive the disempowerment of certain social groups (Lim et al., 2022). For example, students who regularly have access to state-of-the-art AI image generation tools may be better able to distinguish between AI-generated content and human-created content, thus developing a better level of digital literacy and being less likely to fall prey to scams or misinformation. Therefore, a lack of CAIL skills is likely to be an important driver of educational and social inequality.

A further issue in compounding educational inequality from a multiliteracies perspective is the diminishing diversity of voices in educational discourse. GenAI models have been said to produce 'echo chambers' (Turobov et al., 2024) and may lead to overreliance on such systems, which could reinforce existing knowledge hierarchies (Leslie, 2023). Given that the multiliteracies approach focuses on inclusivity and a range of diverse voices, this could damage the potential for educational justice. The biases contained within GenAI image outputs, for example, tend to emphasize and compound stereotypical notions across the lines of gender, skin color, and class (Nicolleti & Bass, 2023). Biases may also be introduced not from the training data but from the developers themselves, who conduct training, monitoring, and adjustments of GenAI model output through supervised learning (Alser & Waisberg, 2023).

Furthermore, GenAI models require vast amounts of resources for their creation and training. The development of physical resources such as Graphics Processing Units (GPUs) is significant (Hosseini et al., 2024), and the production of GenAI images uses a high amount of energy, potentially contributing to climate change (Heikkilä, 2023; Luccioni et al., 2024), which may disproportionately impact countries in the Global South. While technology has created new ways of meaning-making (Lim et al., 2022), it has also introduced complex challenges that threaten the potential for educational inclusion and justice.

### GenAI as a Potential Contributor to a New Multiliteracies

In contrast to these worrying threats to the notion of educational justice, there are arguments that GenAI may contribute to democracy within the educational process, including fostering the development of CAIL and new multiliteracies in the age of GenAI. Research has demonstrated that GenAI tools may be able to support the multiliteracies learning of L2 students (Chandel & Lim, 2024), support student learning in general (Deng et al., 2024), and contribute to the reconceptualization of traditional teaching practices in L2 writing, such as paraphrasing (Roe, O'Sullivan, Arumynathan, et al., 2024). These examples offer new opportunities for developing a multiliteracies-based pedagogy.

Furthermore, there is the potential for GenAI models to be developed in an open-source, ethical framework based on training data that is public and used ethically and consensually. In this case, such forms of more transparent GenAI may be able to assist equitable learning; if all learners can access and benefit from ethical GenAI systems, then there is the potential for these technologies to assist in providing essential learning resources, recommend relevant topics for further study, suggest readings, or give formative feedback. It is even conceivable that such tools can provide structured guidance on how to critically and ethically engage with GenAI. Therefore, GenAI has the potential to benefit multiliteracy development, albeit with the caveat that such models should be deployed equitably and ethically trained.

### Creating New Metaphors to Develop Multiliteracies in the GenAI Era

Our approach to developing CAIL as part of a project in multiliteracies is grounded in Conceptual Metaphor Theory (CMT) (Lakoff & Johnson, 2003), which posits that metaphors are more than just an element of style or a technique that makes language more engaging for readers; metaphors are a significant, pervasive, and powerful tool by which we structure our social worlds and create and recreate realities (Batten, 2012). By metaphor, we draw on Lakoff and Johnson's (2003) assertion that one 'thing' may be explained in terms of another. Prior research has sought to draw attention to the nature of metaphors for describing GenAI and AI systems in general by discussing how the use of metaphor may help learners engage with the otherwise opaque qualities of algorithmic systems. For instance, the use of an echo chamber to represent algorithmic bias or a funhouse mirror to emphasize the ways in which distortions of physical representations may occur (Roe, Furze, & Perkins, 2024). Metaphors have deep significance in relation to literacies, cultures, and social norms. Bialostock (2002, 2008) explored the role of metaphor in literacy development and attitudes towards literacy pedagogy in depth and argued that metaphors may serve as cultural models, which are employed to help members of a society make sense of their world and understand an abstract domain (Bialostok, 2008). In education, the analysis of metaphors for artificial intelligence has been applied to the ways students perceive AI (Demir & Güraksin, 2022), the "colonising" influence of data-driven AI and Generative AI applications (Ferreira et al., 2020; Gupta et al., 2024), and a problematisation of machines' capacity to "think" (Dippel, 2019). Many of these approaches analyze existing metaphors used in the discourse of AI technologies; we propose instead that this new technology requires the creation of *new* metaphors.

In terms of multiliteracies, we propose that metaphors can help us reorganize our knowledge of the world (Saban, 2006). This is significant and in line with the overall goal of multiliteracies to increase educational equity, as we may develop new ways of understanding the implications of unequal access to technology or find new ways to help learners understand abstract concepts (Niemeier, 2017). Research has demonstrated that metaphors of literacy, for example, can become dominant models of understanding for teachers. Consequently, understanding teachers' cultural models of literacies through metaphor analysis can help restructure conversations about literacy pedagogy (S. M. Bialostok, 2014).

We also argue that to engage in the analysis, understanding, and critique of digital products such as GenAI content, learners must develop CAIL. We distinguish CAIL from the broader term AI literacy, which is loosely defined and inconsistently applied, with no universal definition (Bozkurt et al., 2023). We recognize that students need to engage with AI as part of technological multiliteracy (Stolpe & Hallström, 2024). We agree with Long and Magerko (2020) that AI competencies must enable learners to critically evaluate AI technologies and use them effectively for collaboration and communication in multiple contexts, including at home, work, and online.

To support this understanding, we propose a new metaphorical framework that conceptualizes AI-generated content as a form of 'digital plastic'–a substance that, like its physical counterpart, offers both possibilities and challenges to society. Through this metaphorical lens, we argue that learners may better understand how to harness these tools effectively while remaining mindful of their potential impact (Bozkurt et al., 2023).

## Exploring GenAI Content as Digital Plastic

Digital plastic is, like its physical counterpart, synthetic, ubiquitous, helpful and potentially toxic. The metaphor allows for both exploration of the affordances of GenAI and a critique of the potential damage to the online "ecosystem" through which this new media will proliferate. In their discussion of multimodal discourse, Kress and Van Leeuwen (2001) provide a brief history of attitudes towards the manufacture of plastics. Since the 1930s, plastic has been both celebrated and vilified as a medium for production because of its versatility, detrimental impacts on the environment, and mass commercialization of plastic-based products. Barthes (1957) referred to the potential in plastic for "infinite transformation" and its potential as "ubiquity made visible" – both a blessing and a curse, and Baudrillard (1996, p. 112) referred to the "game of the universal associations" made possible through plastics. Applying these same qualities to GenAI, it is possible to view this new "digital plastic" as equally malleable for creating meaning through digital texts.

Like physical plastics, GenAI content may provide new possibilities through its relatively low cost at the point of use. For scholars and writers in the Global South, where access to professional editorial services or native English-speaking collaborators is often restricted by financial and geographical barriers, GenAI tools could help address longstanding inequities in global academic discourse (James & Andrews, 2024). Just as plastic materials made essential goods available to communities with limited resources, GenAI might offer writers from non-Western contexts capabilities that were previously limited to those who could afford expensive language services or had access to Western academic networks. These tools could potentially reduce historical barriers to participation in international publishing and online discourse, particularly for multilingual scholars working in English-medium academic contexts (Kim & Danilina, 2025; Perkins, 2023; Sebastian & Baron, 2024). However, this pattern of adoption mirrors the complex issues seen in physical plastic usage; just as many Global South countries face challenges with plastic waste due to limited regulation and infrastructure, there may be parallel issues with digital plastic. Writers with restricted access to premium GenAI models may rely more heavily on free, potentially lower-quality tools, contributing to a possible increase in lower-quality synthetic content in digital spaces. This pattern suggests the risk of creating new forms of digital inequality, even as it addresses others.

Drawing on Kress (2010), the modal affordances of GenAI include the possible semiotic resources that a user can draw on when using the technology to make meaning. Because GenAI is a multimodal technology, these affordances are many and varied, including the creation of text, images, and audio. The transduction (Kress, 1997) of meaning across modes is also possible, with new technologies capable of taking input in one mode, such as text, and producing novel outputs in another. Advances in GenAI technologies, such as image recognition (OpenAI, 2023) and voice generation (OpenAI, 2024), mean that this potential for the transduction of meaning can occur in various ways, including text-to-image, image-to-speech, and speech-to-text. Similar to physical plastic, GenAI is a malleable and versatile technology. This raises important questions for CAIL, including what is gained or lost when meaning is translated in this way across modes.

For example, although it has been possible for some time to generate compelling and even realistic images from detailed text prompts, Kalantzis and Cope (2024, p.16). argue that "Computers can't mean anything other than zero or one. All they can do is calculate by textual transposition: recorded Unicode > chunked into tokens > binary notation > calculation of the probability of the next token > token > readable Unicode." If this is true, then the seeming affordance of multimodal GenAI may be fundamentally limited by deference to the written mode. If a user describes an image, either verbally or via a text prompt, the image generation technology can only produce a novel image within the constraints of its text-based training data, that is, the labels associated with classified images. In this respect, the AI image generator's mind-eye is more limited than the imagination of a human creator. However, as these models continue to grow in both size and sophistication, and the volume of training data well exceeds the potential for an individual human to view and learn from images, the point may be moot. The implications of this for the future of multimodal content creation should be a priority for future research. The potential implications of AI-generated plastics are similar. Synthetically produced media offers many possibilities for creativity, accessibility, and communication. However, it may bring a flood of low-quality, homogeneous, and culturally biased content into diverse digital ecosystems. The consequences of this may be benign or significantly damaging to those who ingest it. Within this framework, digital plastic can be understood as a new form of multimodal meaning-making that disrupts traditional norms and hierarchies.

Multiliteracies theory emphasizes the diversity of communicative practices and the importance of engaging with multiple modalities. Synthetic media can expand these possibilities by enabling individuals to use GenAI to engage with such productions. For example, a user may be able to produce a project involving AI-generated text narration, visual storytelling, and audio production, all through the use of GenAI tools. However, this democratization of multimodal production also has trade-offs, including the potential to prioritize efficiency over an artistic voice and, therefore, an overreliance on algorithmic production.

Persistence is another shared issue between digital and physical plastics. Physical plastics are enduring and degrade over extended periods, forming microscopic fragments known as microplastics that persist in landfills and oceans for centuries. These degraded particles become permanent pollutants in ecosystems, fundamentally altering the natural environment. This pattern of fragmentation and persistence is parallel to the digital realm of GenAI-generated content. Like their physical counterparts, digital plastics create enduring layers of synthetic content in online environments, with their traces remaining discoverable and measurable through bibliometric studies of journal articles and GenAI footprints (Tang & Eaton, 2024). The concept of digital microplastics manifests when fragments of synthetic content accumulate in the training data for future GenAI models, potentially degrading the quality and reliability of digital ecosystems. This recycling of AI-generated content into training data can lead to model collapse (Shumailov et al., 2024), where the outputs of GenAI models become unintelligible and meaningless, threatening the sustainability of these technologies. Thus, digital plastic mirrors the environmental impact of physical plastics, challenging us to consider how these accumulating layers of synthetic content may reshape the future of meaning-making.

In linking digital plastic to multiliteracies, synthetic media represents both a fundamental shift in how meaning is created and shared in digital spaces and a challenge to our understanding of authenticity and authorship. Multiliteracies theory emphasizes the need for learners to critically engage with diverse communicative practices, and synthetic media pushes these boundaries further, resulting in new ways of thinking about authenticity, authorship, and agency in multimodal contexts. By framing synthetic media as digital plastic, we can better understand the interplay between its materiality and meaning, providing a foundation for addressing its broader implications in education, culture, and society.

## Conclusion

The evolution of GenAI is reshaping much of our extant digital world and may be reaching out into the physical world. Common phrases repeated by GenAI models may soon find greater uptake in public discourse, and GenAI imagery may become an even more common, cost-effective method of producing physical advertisements. As with physical plastics, the integration of synthetic content into daily practices prompts a critical examination of society's growing reliance on artificially generated materials. The scalability, malleability, and ease of use of GenAI suggest that this may soon be the case. However, these same attributes may also lead to toxicity and long-term detriment to societies. The emergence of synthetic media may, from a multiliteracies perspective, disproportionately impact some learners or students more than others, while giving others an outsized benefit in terms of CAIL.

Consequently, we call for new pedagogical strategies, new conceptualizations, and new ways of thinking about GenAI that can help educators raise learners' awareness of these critical properties of AI technologies that may disrupt or damage our digital environments. We argue that the concept of digital plastic can be used as a valuable guiding tool under CMT, through which discussions related to the effects of GenAI can be raised. Practically, we contend that multiliteracies must integrate a component of CAIL as a new skill that must be equitably developed across geographies and cultures to ensure that future generations can identify the effects of digital plastics and microplastics.

**Declaration of Generative AI and AI-assisted technologies in the writing process**

GenAI tools were used for ideation and in some passages of draft text creation which was then heavily revised, along with editing and revision during the production of the manuscript. Tools used were ChatGPT (o3-mini-high) and Claude 3.5 Sonnet, chosen for their ability to provide sophisticated feedback on textual outputs. These tools were selected and used supportively and not to replace core author responsibilities and activities. The authors reviewed, edited, and take responsibility for all outputs of the tools used.